\documentclass [12pt,a4paper]{article}

\title{FORMATION OF COMPLEX MATTER STRUCTURES AND MUTUAL RELATIONS BETWEEN 
THE MASS OF ELEMENTARY PARTICLES}
\author{Jozef \v{S}ima and Miroslav S\'{u}ken\'{\i}k~\\[1ex]
  Slovak Technical University, FCHPT, Radlinsk\'{e}ho 9, \\
  812 37 Bratislava, Slovakia
  \\
  e-mail: sima@chtf.stuba.sk, sukenik@minv.sk}
\date{}
\begin{document}
\maketitle
\begin{abstract}
The model of Expansive Nondecelerative Universe leads to a conclusion 
stating that at the end of radiation era the Jeans mass was equal to the 
upper mass limit of a black hole and, at the same time, the effective 
gravitational range of nucleons was identical to their Compton wavelength. 
At that time nucleons started to exert gravitational impact on their 
environment which enabled to large scale structures become formed. Moreover, 
it is shown that there is a deep relationships between the inertial mass of 
various leptons and bosons and that such relations can be extended also into 
the realm of other kinds of elementary particles. 
\end{abstract}

\section{Introduction}

Particle physics comprises some hundreds of so-called elementary particles. 
Their list is regularly updated, reviewed and published [1]. The particles 
are characterized by their mass (energy), charge, magnetic moment, decay 
mode and mean life, etc. In spite of general believe that there must be very 
deep fundamental interrelationships between the particles parameters, these 
are usually presented and taken into account as independent characteristics. 

In our previous contributions we manifested a coupling of inertial mass for 
some kind of particles, namely for electron, muon, and tau neutrinos [2], 
and for electron, proton, and Planckton [3]. 

This contribution brings the results relating to two aspects of elementary 
particles properties. The first part deals with their recombination leading 
to the formation of large scale structures. The second section represents a 
continuation of our research devoted to unveiling relations between the 
properties of fundamental particles, particularly to the relations between 
the mass of various leptons or bosons.

\section{Recombination and Formation of Large Scale Structures}

Jeans mass $m_{J}$ is the mass at which gravity is balanced by pressure 
forces, i.e.
\begin{equation}
\label{eq1}
{\frac{{G  m_{J}^{2}}}{{r}}} = {\frac{{4\pi}}{{3}}}p r^{3}
\end{equation}
Expressing the mass through the average matter density$\rho $
\begin{equation}
\label{eq2}
m_{J} = {\frac{{4\pi r^{3} \rho}}{{3}}}
\end{equation}
it is obtained for the Jeans mass
\begin{equation}
\label{eq3}
m_{J} = \left( {{\frac{{3}}{{4\pi}} }} \right)^{1 / 2}.{\frac{{p^{3 / 
2}}}{{G^{3 / 2}\rho ^{2}}}}
\end{equation}
where $p$ is the radiation pressure. Except of the relations where numerical 
factors are given by definitions, such factors (e.g. $\left( {3 / 4\pi}  
\right)^{1 / 2}$ in the above equation) will be further omitted. At the end 
of radiation era (the quantities related to this time are denoted by the 
subscript $r)$ the Universe was in the state of thermodynamic equilibrium, 
i.e.
\begin{equation}
\label{eq4}
p_{r} = {\frac{{\rho _{r} c^{2}}}{{3}}}
\end{equation}
The model of Expansive Nondecelerative Universe (further ENU) [4-6] has 
answered the problem of matter density at the end of the radiation era 
giving
\begin{equation}
\label{eq5}
\rho _{r} = {\frac{{3c^{2}}}{{8\pi  G  a_{r}^{2}}} }
\end{equation}
where the gauge factor $a_{r} $ had at the end of the radiation era the 
value of [7]
\begin{equation}
\label{eq:6a}
a_{r} \cong 10^{22}\,\mbox{m}
\end{equation}
Based on (\ref{eq1}) to (6) is follows
\begin{equation}
\label{eq:7a}
m_{J,r} \cong {\frac{{a_{r}  c^{2}}}{{2G}}} \cong 10^{49}\,\mbox{kg}
\end{equation}

Immediately after the recombination, the radiation pressure dropped by $S^{ 
- 1}$ times, where $S$ means the specific entropy defined as the mean number 
of photons per one nucleon. It holds for specific entropy [7,8]
\begin{equation}
\label{eq6}
S \cong 10^{9}
\end{equation}
and thus after the recombination (gauge factor did not significantly changed 
during recombination), $m_{J} $ approached to the value
\begin{equation}
\label{eq:9a}
m_{J} = {\frac{{a_{r}  c^{2}}}{{2G  \left( {S} \right)^{3 / 2}}}} \cong 
10^{35} \,\mbox{kg}
\end{equation}

Stemming from the ENU model background and entropy considerations it was 
possible to estimate an upper mass limit of black holes $m_{(BH)\max}  $ and 
its time evolution [6]. Their gravitational radius $r_{(BH)\max}  $ is 
generally expressed as
\begin{equation}
\label{eq7}
r_{(BH)\max}  = \left( {a^{3}  l_{Pc}}  \right)^{1 / 4}
\end{equation}
At the beginning of the matter era it had to hold
\begin{equation}
\label{eq:11a}
m_{(BH)\max}  = {\frac{{\left( {a_{r}^{3}  l_{Pc}}  \right)^{1 / 4}  
c^{2}}}{{2G}}} \cong 10^{35} \,\mbox{kg}
\end{equation}
which is identical value to that provided by relation~(\ref{eq:9a}).
Putting~(\ref{eq:9a}) and~(\ref{eq:11a}) equal, relation
\begin{equation}
\label{eq8}
S^{6} \cong {\frac{{a_{r}}} {{l_{Pc}}} }
\end{equation}
is obtained. Specific entropy can also be expressed [9] as
\begin{equation}
\label{eq9}
S = {\frac{{m_{p}  c^{2}}}{{h  \nu _{r}}} }
\end{equation}
where $m_{p} $ is the proton mass ($1.67262158\times 10^{ - 27}$ kg) and 
$h\nu _{r} $ is the mean photon energy at the end of the radiation era. From 
the beginning of the Universe expansion up to the end of the radiation era 
the photon energy gradually decreased in time as documented by relation
\begin{equation}
\label{eq10}
h\nu _{r} \approx a^{ - 1 / 2}
\end{equation}

In accordance with (\ref{eq9}) and (\ref{eq10}), the photon energy at
the end of the radiation era was
\begin{equation}
\label{eq11}
h\nu _{r} = m_{Pc}  c^{2}\left( {{\frac{{l_{Pc}}} {{a_{r}}} }} \right)^{1 / 
2}
\end{equation}
where $m_{Pc} $ and $l_{Pc} $ are Planck mass and length [1], respectively
\begin{eqnarray}
\label{eq:16a}
m_{Pc} &=& \left( {{\frac{{\hbar  c}}{{G}}}} \right)^{1 / 2} = 2.176716\times 
10^{ - 8} \,\mbox{kg} \\
\label{eq:17a}
l_{Pc} &=& \left( {{\frac{{G  \hbar}} {{c^{3}}}}} \right)^{1 / 2} = 
1.616051\times 10^{ - 35} \,\mbox{m}
\end{eqnarray}
Then, based on (\ref{eq9}) to~(\ref{eq:17a}) it follows for specific
entropy
\begin{equation}
\label{eq12}
S = {\frac{{m_{p}}} {{m_{Pc}}} }\left( {{\frac{{a_{r}}} {{l_{Pc}}} }} 
\right)^{1 / 2}
\end{equation}
and stemming from (\ref{eq8}) and (\ref{eq12}), at the end of the
radiation era
\begin{equation}
\label{eq13}
a_{r} \cong {\frac{{\hbar ^{2}}}{{G  m_{p}^{3}}} }
\end{equation}
The above relation is of key significance [5,6]. The final relation
for specific entropy, based on~(\ref{eq12}) and~(\ref{eq13}), adopts
the form
\begin{equation}
\label{eq14}
S \cong \left( {{\frac{{m_{Pc}}} {{m_{p}}} }} \right)^{1 / 2}
\end{equation}

\section{Relationships Between the Elementary Particles Masses}

It seems to be obvious that completing the recombination, the Compton
wavelength of nucleons equals to their effective gravitational range.
This is the time when gravitational influence of nucleons on their
environment started to be effective that, in turn, enabled to more
complex matter structures be formed. Before the recombination nucleons
could not exert gravitational impact. The mass in
equation~(\ref{eq:7a}) represents in the ENU the total Universe mass
at the end of radiation era, i.e.\ a limit mass.

The above conclusions suggest the existence of an important
relationship between the mass of nucleons, ionization energy of the
hydrogen atom $E_{(H)} $, and the mass of the electron $m_{e} $. It
should be pointed out that after the recombination, the mass $m_{J} $
is identical to the maximum mass of a black hole at the given time.

The recombination started at the temperature $T_{r} $. Using relation
(\ref{eq14}) it may be written
\begin{equation}
\label{eq15}
S^{ - 1} \cong \left( {{\frac{{m_{p}}} {{m_{Pc}}} }} \right)^{1 / 2} \cong 
\exp \left( { - {\frac{{\Delta E}}{{k  T_{r}}} }} \right)
\end{equation}
where
\begin{equation}
\label{eq16}
\Delta E = E_{(H)} - k  T_{r} 
\end{equation}
Providing that 
\begin{equation}
\label{eq17}
h\nu _{r} \cong k  T_{r} 
\end{equation}
it follows from (\ref{eq15}), (\ref{eq16}), and (\ref{eq17}) that
\begin{equation}
\label{eq18}
h\nu _{r} \cong {\frac{{E_{(H)}}} {{1 - \ln \left( {{\frac{{m_{p}}} {{m_{Pc} 
}}}} \right)^{1 / 2}}}}
\end{equation}
Since the ionization energy of the hydrogen atom can be expressed as
\begin{equation}
\label{eq19}
E_{(H)} \cong {\frac{{\alpha _{e}^{2} m_{e} c^{2}}}{{2}}}
\end{equation}
where $\alpha _{e} $ is the constant of hyperfine structure ($ \cong 
7.3\times 10^{ - 3})$. When (\ref{eq11}) and (\ref{eq18}) are put equal, using (\ref{eq13}) and (\ref{eq19}) 
the relation~(\ref{eq:26a}) relating the electron, proton and Planckton masses is 
obtained 
\begin{equation}
\label{eq:26a}
m_{e} \cong {\frac{{2{\left[ {1 - \ln \left( {{\frac{{m_{p}}} {{m_{Pc}}} }} 
\right)^{1 / 2}} \right]}\left( {{\frac{{m_{p}^{3}}} {{m_{Pc}}} }} 
\right)^{1 / 2}}}{{\alpha _{e}^{2}}} } \cong 4.0\times 10^{ - 31} \,\mbox{kg}
\end{equation}
Taking into account the simplification of some relations (e.g. the
omissions of numerical coefficients), the calculated electron mass is
in good agreement with its known value ($ \cong 9.1\times 10^{ - 31}$
kg).

We believe that relationships analogous to~(\ref{eq:26a}) should exist
also for other couples of particles. In this part we manifest such
connections for some couples of stable leptons and bosons.

Near the energy of 100 GeV the unification of electromagnetic and weak
interaction occurs. The energy corresponds to the mass of vector
bosons Z and W. When the proton mass is substituted in~(\ref{eq:26a})
for the vector boson W mass, and adjusted value [10] of the splitting
constant is taken into calculation, relation~(\ref{eq:26a}) leads to
the lepton $\mu $ mass ($\sim $ 230 $m_{e} $) which is very close to
its actual mass $(206.7m_{e})$. Another relation can be found for the
heavy lepton $\tau $ mass (which is about $3.03\times 10^{ - 27}$ kg
[1]) and the mass of one of the Higgs bosons substituting the proton
mass in~(\ref{eq:26a}) for the Higgs boson mass $7\times 10^{ - 25}$ kg [1].

It may be demonstrated that~(\ref{eq:26a}) can be taken as a bridge
between the macro-world (the Universe) and the micro-world
(particles). Substitution of the proton mass in~(\ref{eq:26a}) by the
value of $5.35\times 10^{ - 12}$ kg (the mass of bosons X,Y) leads
directly to Planck mass.

\section*{References}

\begin{description}

\item {}[1] D.E. Groom et al., Eur. Phys. J., C15 (2000) 1
\item {}[2] M. S\'{u}ken\'{\i}k, J. \v{S}ima, gr-qc/0012044
\item {}[3] J. \v{S}ima, M. S\'{u}ken\'{\i}k, gr-qc/0011057
\item {}[4] V. Skalsk\'{y}, M. S\'{u}ken\'{\i}k, Astrophys. Space
  Sci., 178 (1991) 169
\item {}[5] V. Skalsk\'{y}, M. S\'{u}ken\'{\i}k, Astrophys. Space
  Sci., 181 (1991) 153
\item {}[6] J. \v{S}ima, M. S\'{u}ken\'{\i}k, Spacetime {\&} Substance
  2 (2001) 79
\item {}[7] M. S\'{u}ken\'{\i}k, J. \v{S}ima, gr-qc/0106078
\item {}[8] M. S\'{u}ken\'{\i}k, J. \v{S}ima, Entropy, submitted
\item {}[9] M. S\'{u}ken\'{\i}k, J. \v{S}ima, gr-qc/0103028
\item {}[10] N.N. Bogolyubov, D.V. Shirkov, Adv. Math. Phys. Astronomy
  (in Czech), 32 (1987) 251
\end{description}

\end{document}